\documentclass[12pt]{article}

\usepackage{amsmath}
\usepackage{amssymb}
\usepackage[dvips]{graphicx}
\usepackage{setspace}
\usepackage{epsfig}

\newcommand{\K}{{\sf{K}}}

\newcommand{\Y}{\mathbf{Y}}

\newcommand{\R}{{\mathbf{R}}}

\newcommand{\C}{{\sf{C}}}

\newcommand{\tsnr}{{\text{\footnotesize{SNR}}}}
\newcommand{\ssnr}{\text{\scriptsize{SNR}}}
\newcommand{\ud}{\mathrm{d}}

\newtheorem{prop:coherentoofskcap}{Proposition}
\newtheorem{prop:oofskcap}[prop:coherentoofskcap]{Proposition}
\newtheorem{cor:coherentcapMinf}[prop:coherentoofskcap]{Proposition}
\newtheorem{cor:capMinf}[prop:coherentoofskcap]{Proposition}

\newtheorem{cor:finiteParSlope}[prop:coherentoofskcap]{Proposition}

\newtheorem{cor:fixedPeakLimit}[prop:coherentoofskcap]{Proposition}

\newtheorem{prop:coherentoofskcapwithphase}[prop:coherentoofskcap]{Proposition}
\newtheorem{prop:oofskcapwithphase}[prop:coherentoofskcap]{Proposition}
\newtheorem{cor:coherentcapMinfwithphase}[prop:coherentoofskcap]{Proposition}
\newtheorem{cor:capMinfwithphase}[prop:coherentoofskcap]{Proposition}

\newtheorem{cor:finiteParSlopewithphase}[prop:coherentoofskcap]{Proposition}

\newtheorem{cor:fixedPeakLimitwithphase}[prop:coherentoofskcap]{Proposition}

\newtheorem{lemma:martingale}{Lemma}

\newtheorem{rem:cohnoncohloss}{Remark}
\newtheorem{rem:Minfasymptotics}[rem:cohnoncohloss]{Remark}
\newtheorem{rem:cohnoncohlosswithphase}[rem:cohnoncohloss]{Remark}
\newtheorem{rem:Minfasymptoticswithphase}[rem:cohnoncohloss]{Remark}

\begin{document}

\title{On-Off Frequency-Shift-Keying for Wideband Fading Channels \footnote{This paper was prepared through collaborative
participation in the Communications and Networks Consortium
sponsored by the U.S. Army Research Laboratory under the
Collaborative Technology Alliance Program, Cooperative Agreement
DAAD19-01-2-0011. The U.S. Government is authorized to reproduce
and distribute reprints for Government purposes notwithstanding
any copyright notation thereon. The material in this paper was
presented in part at the IEEE Vehicular Technology Conference
(VTC), Los Angeles, CA, Sept. 26-29 2004, and the IEEE Global
Telecommunications Conference (Globecom), Dallas, TX, 29 Nov. - 3
Dec. 2004.}}

\author{Mustafa Cenk Gursoy \footnote{Mustafa Cenk Gursoy was with the Department of Electrical Engineering,
Princeton University, Princeton, NJ 08544; he is now with the
Department of Electrical Engineering, University of
Nebraska-Lincoln, Lincoln, NE 68588 (e-mail :
gursoy@engr.unl.edu).} \and H. Vincent Poor $^\ddagger$ \and
Sergio Verd\'u \footnote{H. Vincent Poor and Sergio Verd\'u are
with the Department of Electrical Engineering, Princeton
University, Princeton, NJ 08544 (e-mail : poor@princeton.edu;
verdu@princeton.edu).}}

\date{}

\maketitle

\begin{spacing}{1.6}

\begin{abstract}
M-ary On-Off Frequency-Shift-Keying (OOFSK) is a digital modulation
format in which M-ary FSK signaling is overlaid on On/Off keying.
This paper investigates the potential of this modulation format in
the context of wideband fading channels. First it is assumed that
the receiver uses energy detection for the reception of OOFSK
signals. Capacity expressions are obtained for the cases in which
the receiver has perfect and imperfect fading side information.
Power efficiency is investigated when the transmitter is subject to
a peak-to-average power ratio (PAR) limitation or a peak power
limitation. It is shown that under a PAR limitation, it is extremely
power inefficient to operate in the very low $\tsnr$ regime. On the
other hand, if there is only a peak power limitation, it is
demonstrated that power efficiency improves as one operates with
smaller $\tsnr$ and vanishing duty factor. Also studied are the
capacity improvements that accrue when the receiver can track phase
shifts in the channel or if the received signal has a specular
component. To take advantage of those features, the phase of the
modulation is also allowed to carry information.
\\
\noindent \emph{Keywords:} Frequency-shift keying, On-Off keying,
fading channels, Rician fading, channel capacity, power
efficiency, peak constraints, wideband regime.

\end{abstract}

\section{Introduction} \label{sec:oofskintro}

A wide range of digital communication systems in wireless,
deep-space and sensor networks operate in the low-power regime
where power consumption rather than bandwidth is the limiting
factor. For such systems, power-efficient transmission schemes are
required for effective use of scarce energy resources. For
example, in sensor networks \cite{Akyildiz}, nodes that are
densely deployed in a region may be equipped with only a limited
power source and in some cases replenishment of these resources
may not be possible. Therefore, energy-efficient operation is
vital in these systems. Recently there has also been much interest
in ultrawideband systems in which low-power pulses of very short
duration are used for communication over short distances. These
wideband pulses must satisfy strict peak power requirements in
order not to interfere with existing
systems. 

The power efficiency of a communication system can be measured by
the energy required for reliable communication of one bit. When
communicating at rate $R$ bits/s with power $P$, the transmitted
energy per bit is $E_b = \frac{P}{R}$. Since the maximum rate is
given by the channel capacity, $C$, the least amount of bit energy
required for reliable communication is $E_b = \frac{P}{C}$. In
\cite{Shannon}, Shannon showed that the capacity of an ideal
bandlimited additive white Gaussian noise channel is $C =
B\log_2\left(1 + \frac{P}{BN_0} \right)$ bits/s where $P$ is the
received power, $B$ is the channel bandwidth and $N_0$ is the
one-sided noise spectral level. As the bandwidth grows to infinity,
the capacity monotonically increases to $\frac{P}{N_0} \log_2e$
bits/s, therefore decreasing the required received bit-energy
normalized to the noise power to
\begin{gather}
\frac{E_b^r}{N_0} = \frac{P/N_0}{C} \underset{B \to
\infty}{\longrightarrow} \log_e2 = -1.59 \text{ dB}.
\label{eq:minbitenergy}
\end{gather}
This minimum bit energy (\ref{eq:minbitenergy}) can be approached by
pulse position modulation with vanishing duty cycle \cite{Golay} or
by $M$-ary orthogonal signaling as $M$ becomes large \cite{Turin}.
In the presence of unknown fading, Jacobs \cite{Jacobs} and Pierce
\cite{Pierce} have noted that $M$-ary orthogonal signaling obtained
by frequency shift keying (FSK) modulation can still approach the
limit in (\ref{eq:minbitenergy}) for large values of $M$. Gallager
\cite[Sec. 8.6]{Gallager} also demonstrated that over fading
channels $M$-ary orthogonal FSK signaling with vanishing duty cycle
approaches  the infinite bandwidth capacity of unfaded Gaussian
channels as $M \to \infty$, thereby achieving
(\ref{eq:minbitenergy}). The result that the infinite bandwidth
capacity of fading channels is the same as that of unfaded Gaussian
channels is also noted by Kennedy \cite{Kennedy}. Telatar and Tse
\cite{Telatar Tse} considered a more ganeral fading channel model
that consists of a finite number of time-varying paths and showed
that the infinite bandwidth capacity of this channel is again
approached by using peaky FSK signaling.  Luo and M\'edard
\cite{LuoMedard} have shown that FSK with small duty cycle can
achieve rates of the order of capacity in ultrawideband systems with
limits on bandwidth and peak power. Reference \cite{Verdu} shows, in
wider generality than was previously known, that the minimum
received bit energy normalized to the noise level in a Gaussian
channel is $-1.59$ dB regardless of the knowledge of the fading at
the receiver and/or transmitter. It is also shown in \cite{Verdu}
that if the receiver does not have perfect knowledge of the fading,
flash signaling is required to achieve the minimum bit energy. The
performance degradation in the wideband regime incurred by using
signals with limited peakedness is discussed in \cite{Medard
Gallager}, \cite{Telatar Tse}, and \cite{gursoy2}. The error
performance of FSK signals used with a duty cycle is analyzed in
\cite{Lun} and \cite{LuoMedard2}.

Besides approaching the minimum energy per bit, FSK modulation is
particularly suitable for noncoherent communications. Butman
\emph{et al.} \cite{Butman Bar-David} studied the performance of
$M$-ary FSK, which has unit peak-to-average power ratio, over
noncoherent Gaussian channels by computing the capacity and
computational cutoff rate.  Stark \cite{Stark} analyzed the
capacity and cutoff rate of $M$-ary FSK signaling with both hard
and soft decisions in the presence of Rician fading and noted that
there exists an optimal code rate for which the required bit
energy is minimized.

In this paper, we study the power efficiency of $M$-ary On/Off FSK
(OOFSK) signaling in which $M$-ary FSK signaling is overlaid on
top of On/Off keying, enabling us to introduce peakedness in both
time and frequency. Our main focus will be on cases in which the
peakedness of input signals is limited. The organization of the
paper is as follows. Section \ref{sec:oofskmodel} introduces the
channel model. In Section \ref{sec:oofsk}, we find the capacity of
$M$-ary orthogonal OOFSK signaling with energy detection at the
receiver and investigate the power efficiency in two cases:
limited peak-to-average power ratio and limited peak power. In
Section \ref{sec:oofskwithphase}, we consider joint frequency and
phase modulation and analyze the capacity and power efficiency of
$M$-ary OOFPSK signaling in which the phase of FSK signals also
convey information. Finally, Section \ref{sec:oofskconclusion}
includes our conclusions.


\section{Channel Model} \label{sec:oofskmodel}

In this section, we present the system model. We assume that
$M$-ary orthogonal OOFSK signaling, in which FSK signaling is
combined with On-Off keying with a fixed duty factor, $\nu \le 1$,
is employed at the transmitter for communication over a fading
channel. In this signaling scheme, over the time interval of
$[0,T]$ the transmitter either sends no signal with probability
$1-\nu$ or sends one of $M$ orthogonal sinusoidal signals,
\begin{gather}
s_i(t) = \sqrt{\frac{P}{\nu}} \, \, e^{j(\omega_it + \theta_i)}
\quad 0 \le t \le T, \quad 1 \le i \le M,
\end{gather}
with probability $\nu$. To ensure orthogonality, adjacent
frequency slots satisfy $|\omega_{i+1}-\omega_i| =
\frac{2\pi}{T}$. Choosing $\nu = 1$, we obtain ordinary FSK
signaling. If the channel input is $X = i$ for $1 \le i \le M $,
the transmitter sends the sine wave $s_i(t)$, while no
transmission is denoted by $X = 0$. Note that OOFSK signaling has
average power $P$, and peak power $P/\nu$. We assume that the
transmitted signal undergoes stationary and ergodic fading and
that the delay spread of the fading is much less than the symbol
duration. Under these assumptions, the fading has a multiplicative
effect on the transmitted signal and the received signal can be
modeled as follows:
\begin{gather}
r(t) = h(t)\, s_{X_k}(t - (k-1)T) + n(t), \quad (k-1)T \le t \le kT,
\quad \text{ for }k = 1,2, \ldots, \nonumber
\end{gather}
where $\{X_k\}_{k = 1}^\infty$ is the input sequence with $X_k \in
\{0 ,1, 2, \ldots, M\}$, $h(t)$ is a proper\footnote {See
\cite{Neeser}.} complex stationary ergodic fading process with
$E\{h(t)\} = d$ and $\text{var}(h(t)) =\gamma^2$, and $n(t)$ is a
zero-mean circularly symmetric complex white Gaussian noise
process with single-sided spectral density $N_0$. Note that
$s_0(t) = 0$. If we further assume that the symbol duration $T$ is
less than the coherence time of the fading, then the fading stays
constant over the symbol duration and the channel model now
becomes
\begin{gather}
r(t) = h_k \, s_{X_k}(t - (k-1)T) + n(t), \quad (k-1)T \le t \le
kT. \label{eq:oofskmodel}
\end{gather}
At the receiver, a bank of correlators is employed in each symbol
interval to obtain the $M$-dimensional vector $\Y_k = (Y_{k,1},
\ldots, Y_{k,M})$ where
\begin{gather}
Y_{k,i} = \frac{1}{\sqrt{N_0T}} \int_{(k-1)T}^{kT} r(t)
e^{-j\omega_it} \,\ud t, \quad i = 1,2, \ldots, M. \label{eq:yk}
\end{gather}
It is easily seen that, given the symbol $X_k = i$, phase
$\theta_{i}$ and fading coefficient $h_k$, $Y_{k,j}$ is a proper
complex Gaussian random variable with
\begin{align}
E\{Y_{k,j} | X_k = i, \theta_{i}, h_k\} = \alpha \, h_k \, e^{j
\theta_{i}} \delta_{ij} \quad \text{and} \quad \text{var}(Y_{k,j} |
X_k = i, \theta_{i}, h_k) = 1, \nonumber
\end{align}
where $\delta_{ij} = 1$ if $i = j$ and is zero otherwise, and
$\alpha^2 = \frac{PT}{\nu N_0} = \frac{\ssnr}{\nu}$ with $\tsnr$
denoting the signal-to-noise ratio per symbol.


\section{Capacity of $M$-ary Orthogonal OOFSK Signaling with Energy Detection} \label{sec:oofsk}

In this section, we analyze the capacity of $M$-ary orthogonal
OOFSK signaling when in every symbol interval, the noncoherent
receiver measures the energy at each of the $M$ frequencies, i.e.,
computes
\begin{gather}
R_{k,i} = |Y_{k,i}|^2 = \left| \frac{1}{\sqrt{N_0T}}
\int_{(k-1)T}^{kT} r(t) e^{-j\omega_it} \,\ud t \right|^2, \quad 1
\le i \le M, \quad \text{for } k = 1,2,\ldots,
\end{gather}
and the decoder sees the vector $\R_k = (R_{k,1}, \ldots, R_{k,M})$.
With this structure, the receiver does not need to track phase
changes in the channel. We consider the cases where the receiver has
either perfect or imperfect fading side information while the
transmitter has no knowledge of the fading coefficients. Besides
providing the ultimate limits on the rate of communication, capacity
results also offer insight into the power efficiency of OOFSK
signaling by enabling us to obtain the energy required to send one
bit of information reliably.

In the low-power regime, the spectral-efficiency/bit-energy tradeoff
reflects the fundamental tradeoff between bandwidth and power.
Assuming that the bandwidth of $M$-ary OOFSK modulation is
$\frac{M}{T}$ where $T$ is the symbol duration, the maximum
achievable spectral efficiency is
\begin{gather}
\C\left( \frac{E_b}{N_0} \right) = \frac{1}{M} \,\, C(\ssnr) \quad
\text{bits/s/Hz} \label{eq:specteff}
\end{gather}
where $C(\ssnr)$ is the capacity in bits/symbol, and
\begin{gather}
\frac{E_b}{N_0} = \frac{\ssnr}{C(\ssnr)}
\end{gather}
is the bit energy normalized to the noise power. For average power
limited channels, the bit energy required for reliable
communications decreases monotonically with decreasing spectral
efficiency, and the minimum bit energy is achieved at zero spectral
efficiency, i.e., $\frac{E_b}{N_0}_{\min} = \lim_{\ssnr \to 0}
\frac{\ssnr}{C(\ssnr)} = \frac{\log_e2}{\dot{C}(0)}$ where
$\dot{C}(0)$ is the first derivative of the capacity in nats. Hence
for fixed rate transmission, reduction in the required power comes
only at the expense of increased bandwidth. Reference [7] analyzes
the spectral-efficiency/bit-energy function in the low-power regime
for a general class of average power limited fading channels and
shows that the minimum bit energy is $\log_e2 = -1.59 \,\,
\text{dB}$ as long as the additive background noise is Gaussian.
This minimum bit energy is achieved only in the asymptotic regime of
infinite bandwidth. If one is willing to spend more power, then
reliable communication over a finite bandwidth is possible. Hence
achieving the minimum bit energy is not a sufficient criterion for
finite bandwidth analysis. The wideband slope \cite{Verdu}, defined
as the slope of the spectral efficiency curve
${\sf{C}}\left(\frac{E_b}{N_0}\right)$ in bits/s/Hz/3dB at zero
spectral efficiency, is given by:
\begin{align}
S_0 &\stackrel{\textrm{def}}{=} \lim_{\frac{E_b}{N_0} \downarrow
\left.\frac{E_b}{N_0}\right|_{{\sf{C}} = 0}} \frac{{\sf{C}}\left(
\frac{E_b}{N_0}\right)}{10\log_{10}\frac{E_b}{N_0} - 10\log_{10}
\left. \frac{E_b}{N_0} \right|_{{\sf{C}} = 0}} \,\,10 \log_{10}2
\nonumber \\ &=
\frac{1}{M}\frac{2\left(\dot{C}(0)\right)^2}{-\ddot{C}(0)},
\label{eq:widebandslope}
\end{align}
where $\dot{C}(0)$ and $\ddot{C}(0)$ denote the first and second
derivatives of the capacity in nats. Note that differing from the
original definition in \cite{Verdu}, normalization by $M$ is
introduced in (\ref{eq:widebandslope}) due to the scaling in
(\ref{eq:specteff}). The wideband slope closely approximates the
growth of the spectral efficiency curve in the power-limited
regime and hence is a useful tool providing insightful results
when bandwidth is a resource to be conserved.


\subsection{Perfect Receiver Side Information}

We first assume that the receiver has perfect knowledge of the
magnitude of the fading, $|h|$. For this case, the capacity as a
function of $\ssnr = \frac{PT}{N_0}$ of $M$-ary OOFSK signaling
with energy detection is given by the following proposition.
Throughout the paper, we denote the probability density function
and distribution function of a random variable $Z$ by $p_Z$ and
$F_{Z}$, respectively, with arguments omitted in equations in
order to avoid cumbersome expressions.

\begin{prop:coherentoofskcap} \label{prop:coherentoofskcap}
Consider the fading channel model (\ref{eq:oofskmodel}) and assume
that the receiver knows the magnitude but not the phase of the
fading coefficients $\{h_k, k = 1,2,\ldots\}$. Further assume that
the transmitter has no fading side information. Then the capacity
of $M$-ary orthogonal OOFSK signaling with a fixed duty factor
$\nu \le 1$ with energy detection is
\begin{align}
C^p_M(\ssnr) = E_{|h|} \left\{(1-\nu)\int p_{\R |X = 0} \log
\frac{p_{\R | X = 0}}{p_{\R | \, |h|}} \, \ud \R + \nu\int p_{\R
|X = 1, |h|} \log \frac{p_{\R | X = 1, |h| }}{p_{\R | \, |h|}} \,
\ud \R \right\} \label{eq:coherentoofskcap}
\end{align}
where
\begin{gather}
p_{\R | \, |h|} = (1-\nu)p_{\R |X=0} + \frac{\nu}{M} \sum_{i=1}^M
p_{\R |X=i, |h|}, \label{eq:p(R)}
\\
p_{\R|X=0} = e^{-\sum_{j=1}^M R_j }, \label{eq:p(R|X=0)}
\\
p_{\R |X=i, |h|} = e^{-\sum_{j=1}^M R_j} f(R_i,|h|,\ssnr) \quad 1
\le i \le M, \label{eq:p(R|X = i)}
\\
\intertext{and} f(R_i,|h|,\ssnr) = \exp\left( -\ssnr/\nu \,
|h|^2\right) I_0\left( 2\sqrt{\ssnr/\nu \, |h|^2 R_i}\right).
\label{eq:fRi}
\end{gather}
\end{prop:coherentoofskcap}
\noindent \textbf{Proof}: See Appendix
\ref{appendix:coherentoofskcap}.

Formula (\ref{eq:coherentoofskcap}) must be evaluated numerically,
and computational complexity imposes a burden on numerical
techniques for large $M$. Fortunately, a simpler expression is
obtained in the limit $M \rightarrow \infty$.

\begin{cor:coherentcapMinf} \label{cor:coherentcapMinf}
The capacity expression (\ref{eq:coherentoofskcap}) for $M$-ary
OOFSK signaling in the limit as $M \uparrow \infty$ becomes
\begin{gather}
C_{\infty}^p(\ssnr) = D(p_{R |\,\tilde{x}, |h|} \, \big\| \, p_{R|
\, \tilde{x} = 0, |h| } \big| \, F_{|h|} F_{\tilde{x}})
\end{gather}
where $$R = |y|^2 = |h\tilde{x} + n|^2,$$ $\tilde{x}$ is a
two-mass-point discrete random variable with the following
mass-point locations and probabilities,
\begin{gather}
\tilde{x} = \left\{
\begin{array}{ll}
0 & \text{w.p. } 1-\nu
\\
\sqrt{\frac{\ssnr}{\nu}} & \text{w.p. } \nu,
\end{array} \right. \label{eq:tildex}
\end{gather}
and $n$ is zero-mean circularly symmetric complex Gaussian random
variable with $E\{|n|^2\} = 1$. Therefore,
\begin{align*}
p_{R | \tilde{x}, |h| } &= e^{-R - \tilde{x}^2|h|^2} I_0\left(
2\sqrt{\tilde{x}^2 \, |h|^2 R}\right).
\end{align*}
\end{cor:coherentcapMinf}
\noindent \textbf{Proof}: See Appendix
\ref{appendix:coherentcapMinf}.


\subsection{Imperfect Receiver Side Information} \label{subsec:imperfect}

In this section, we assume that neither the receiver nor the
transmitter has any side information about the fading. Unlike the
previous section, here we consider a more special fading process:
memoryless Rician fading where each of the i.i.d. $h_k$'s is a
proper complex Gaussian random variable with $E\{h_k\} = d$ and
$\text{var}(h_k) = \gamma^2$. Note that the unknown Rician fading
channel can also be regarded as an imperfectly known fading channel
where the specular component is the channel estimate and the fading
component is the Gaussian-distributed error in the estimate. As
argued in \cite{AbouFaycalMedard}, the Bayesian least-squares
estimation over the Rayleigh channel leads to such a channel model.
However, we want to emphasize that no explicit channel estimation
method is considered in this section.

The following result gives the maximum rate at which reliable
communication is possible with OOFSK signaling using energy
detection over the memoryless Rician fading channel. As noted in
Section \ref{sec:oofskintro}, the capacity of the special case of
$M$-ary FSK signaling ($\nu = 1$) was previously obtained by Stark
\cite{Stark}.

\begin{prop:oofskcap}
Consider the fading channel (\ref{eq:oofskmodel}) and assume that
the fading process $\{h_k\}$ is a sequence of i.i.d. proper
complex Gaussian random variables with $E\{h_k\} = d$ and
$\text{var}(h_k) = \gamma^2$ which are not known at either the
receiver or the transmitter. Further assume that energy detection
is performed at the receiver. Then the capacity of $M$-ary
orthogonal OOFSK signaling with fixed duty factor $\nu \le 1$ is
given by
\begin{gather} \label{eq:oofskcap}
C_M^{ip}(\ssnr) = (1-\nu) \int p_{\R|X = 0} \log
\frac{p_{\R|X=0}}{p_{\R}} \, \ud \R + \nu \int p_{\R|X = 1} \log
\frac{p_{\R|X=1}}{p_{\R}} \, \ud \R
\end{gather}
where
\begin{gather}
p_{\R} = (1-\nu)p_{\R|X=0} + \frac{\nu}{M} \sum_{i=1}^M
p_{\R|X=i}, \label{eq:p(R)imperfect}
\\
p_{\R|X=0} = e^{-\sum_{j=1}^M R_j }, \label{eq:p(R|X=0)imperfect}
\\
p_{\R|X=i} = e^{-\sum_{j=1}^M R_j} f(R_i,\tsnr) \quad 1 \le i \le
M, \label{eq:p(R|X=i)imperfect}
\\
\intertext{and} f(R_i,\tsnr) = \frac{1}{\gamma^2 \ssnr/\nu + 1}
\exp\left( \frac{\ssnr/\nu(\gamma^2 R_i - |d|^2 )}{\gamma^2
\ssnr/\nu + 1}\right) I_0\left( \frac{2\sqrt{\ssnr/\nu \, |d|^2
R_i}}{\gamma^2 \ssnr/\nu + 1}\right). \label{eq:fRiimperfect}
\end{gather}
\end{prop:oofskcap}
\noindent \textbf{Proof}: With the memoryless assumption, the
capacity of the $M$-ary OOFSK signaling can be formulated as the
maximum mutual information between the channel input $X_k$ and
output vector $\R_k$ for any $k$. Thus, considering a generic
symbol interval, and dropping the time index $k$, we have
\begin{align*} C &= \max_X \, I(X;\R)
\\
&= \max_X \, (1-\nu) \int p_{\R|X = 0} \log
\frac{p_{\R|X=0}}{p_{\R}} \, \ud \R + \sum_{i=1}^M P(X = i) \int
p_{\R|X = i} \log \frac{p_{\R|X=i}}{p_{\R}} \, \ud \R.
\end{align*}
Similarly as in the proof of Proposition
\ref{prop:coherentoofskcap}, due to the symmetry of the channel, an
input distribution equiprobable over nonzero input values, i.e.,
$P(X = i) = \frac{\nu}{M}$ for $1 \le i \le M$ where $P(X=0)=1-\nu$
achieves the capacity and we easily obtain (\ref{eq:oofskcap}) by
noting that conditioned on $X = i$, $R_j = |Y_j|^2$ is a chi-square
random variable with two degrees of freedom, or more generally,
\begin{gather*}
p_{R_j| X = i} = \left\{
\begin{array}{ll}
\frac{1}{\alpha^2 \gamma^2 + 1} \exp\left( -\frac{R_j + \alpha^2
|d|^2 }{\alpha^2 \gamma^2 + 1}\right) I_0 \left( \frac{2
\sqrt{\alpha^2 |d|^2 R_j} }{\alpha^2 \gamma^2 + 1 }\right) & j = i
\\
e^{-R_j} & j \neq i
\end{array} \right.
\end{gather*}
where, as before, $\alpha^2 = \frac{P T}{\nu N_0}$. Note also that
due to the orthogonality of signaling the vector $\R$ has
independent components and we denote $\tsnr = \frac{PT}{N_0}$.
\hfill $\square$

Similarly to Proposition \ref{cor:coherentcapMinf}, we can find the
infinite bandwidth capacity achieved as the number of orthogonal
frequencies increases without bound. The proof is omitted as it
follows along the same lines as in the proof of Proposition
\ref{cor:coherentcapMinf}.
\begin{cor:capMinf}
The capacity expression (\ref{eq:oofskcap}) of $M$-ary OOFSK
signaling in the limit as $M \uparrow \infty$ becomes
\begin{gather}
C_{\infty}^{ip}(\ssnr) = D(p_{R|\tilde{x}} \, \big\| \, p_{R |
\tilde{x}=0 } \big| F_{\tilde{x}})
\end{gather}
where $$R = |y|^2 = |h\tilde{x} + n|^2,$$  $\tilde{x}$ is a
two-mass-point discrete random variable with mass-point locations
and probabilities given in (\ref{eq:tildex}), and $n$ is a
zero-mean circularly symmetric complex Gaussian random variable
with $E\{|n|^2\} = 1$. Therefore,
\begin{align*}
p_{R | \tilde{x}} &= \frac{1}{\gamma^2 \tilde{x}^2 + 1} \exp\left(
-\frac{R + \tilde{x}^2 |d|^2 }{\gamma^2 \tilde{x}^2 + 1}\right)
I_0 \left( \frac{2 \sqrt{\tilde{x}^2 |d|^2 R} }{\gamma^2
\tilde{x}^2 + 1 }\right).
\end{align*}
\end{cor:capMinf}

\noindent The following remarks are given for the asymptotic case in
which $M$ grows to infinity.

\begin{rem:cohnoncohloss}
\emph{Assume that in the case of perfect receiver side
information, $\{h_k\}$ is a sequence of i.i.d. proper complex
Gaussian random variables. Then the asymptotic loss in capacity
incurred by not knowing the fading is
\begin{align}
C_{\infty}^p(\ssnr) - C_{\infty}^{ip}(\ssnr) &= D(p_{R
|\,\tilde{x}, |h|} \, \big\| \, p_{R| \, \tilde{x} = 0, |h| }
\big| \, p_{|h|} P_{\tilde{x}}) -  D(p_{R|\tilde{x}} \, \big\| \,
p_{R | \tilde{x}=0 } \big| P_{\tilde{x}}) \nonumber
\\
& = I(|h| ; R \, \big| \, \tilde{x})
\end{align}
where $R = |h\tilde{x} + n|^2$.}
\end{rem:cohnoncohloss}

\begin{rem:Minfasymptotics}
\emph{Consider the case of imperfect receiver side information
where
\begin{align}
&C_{\infty}^{ip} = D(p_{R|\tilde{x}} \, \big\| \, p_{R |
\tilde{x}=0 } \big| P_{\tilde{x}}) \nonumber \\ &= (\gamma^2 +
|d|^2) \ssnr - \nu \log\left( \gamma^2 \frac{\ssnr}{\nu} + 1
\right) - \frac{2\ssnr |d|^2}{\gamma^2 \ssnr/\nu+1} + \nu
E_R\left\{\log I_0 \left( \frac{2 \sqrt{\frac{\ssnr}{\nu} |d|^2 R}
}{\gamma^2 \frac{\ssnr}{\nu} + 1} \right) \right\}
\label{eq:Minfnoncohexplicit}
\end{align}
with $\ssnr = \frac{PT}{N_0}$. From (\ref{eq:Minfnoncohexplicit})
we can easily see that for fixed symbol interval $T$,
\begin{gather}\label{eq:Minfnoncohv}
\lim_{\nu \downarrow 0} \frac{1}{T} \, C_{\infty}^{ip}(\ssnr) =
\frac{1}{T}(\gamma^2 + |d|^2) \ssnr = (\gamma^2 + |d|^2)
\frac{P}{N_0} \text{ nats/s},
\end{gather}
and for fixed duty factor $\nu$,
\begin{gather}\label{eq:MinfnoncohT}
\lim_{T \uparrow \infty} \frac{1}{T} \, C_{\infty}^{ip}(\ssnr) =
(\gamma^2 + |d|^2) \frac{P}{N_0} \text{ nats/s}.
\end{gather}
Note that right-hand sides of (\ref{eq:Minfnoncohv}) and
(\ref{eq:MinfnoncohT}) are equal to the infinite bandwidth
capacity of the unfaded Gaussian channel with the same received
signal power. Hence, these results agree with previous results
\cite{Jacobs}, \cite{Pierce} and \cite{Gallager} where it has been
shown that the capacity of $M$-ary FSK signaling over noncoherent
fading channels approaches the infinite bandwidth capacity of the
unfaded Gaussian channel for large $M$ and large symbol duration
$T$ or small duty factor $\nu$.}
\end{rem:Minfasymptotics}


\subsection{Limited Peak-to-Average Power Ratio} \label{subsec:limitedPAR}

The peak-to-average power ratio (PAR) of OOFSK signaling is equal
to the inverse of the duty factor, $1/\nu$. In this section, we
examine the low-{\tsnr} behavior when we keep the duty factor
fixed while the average power $P$ vanishes. We show that under
this limited PAR condition, OOFSK communication with energy
detection at low {\tsnr } values is extremely power inefficient
even in the unfaded Gaussian channel.
\begin{cor:finiteParSlope} \label{prop:finitePARslope}
The first derivative of the capacity at zero {\tsnr} achieved by
$M$-ary OOFSK signaling with a fixed duty factor $\nu \le 1$ over
the unfaded Gaussian channel is zero, i.e., $\dot{C}_M^g(0) = 0$ and
hence the bit energy required at zero spectral efficiency is
infinite,
\begin{gather}
\left.\frac{E_b}{N_0}\right|_{\C = 0} = \lim_{\ssnr \to 0}
\frac{\ssnr}{C_M^g(\ssnr)} \log_e2 = \frac{\log_e2}{\dot{C}_M^g(0)}
= \infty.
\end{gather}
\end{cor:finiteParSlope}
\noindent \textbf{Proof}: Since we consider the unfaded Gaussian
channel, we set the fading variance $\gamma^2 = 0$ in the capacity
expression (\ref{eq:oofskcap}). Note that the only term in
(\ref{eq:oofskcap}) that depends on the signal to noise ratio is
$f(R_i, \ssnr) = \exp(-|d|^2 \ssnr) I_0(2\sqrt{\ssnr |d|^2 R_i})$
in (\ref{eq:fRiimperfect}). Using the fact that $\lim_{x \to 0}
\frac{I_1(a\sqrt{x})}{\sqrt{x}} = \frac{a}{2}$ for $a \ge 0$, one
can show that the derivative at $\tsnr = 0$ is $\dot{f}(R_i,0) =
|d|^2 (-1 + R_i)$. The result then follows by taking the
derivative of the capacity (\ref{eq:oofskcap}) and evaluating it
at $\tsnr = 0$. \hfill $\square$

Since the presence of fading that is unknown at the transmitter does
not increase the capacity, from Proposition
\ref{prop:finitePARslope}, we immediately conclude that $\dot{C}(0)
= 0$ for fading channels regardless of receiver side information as
long as $\nu$ is fixed and hence the peak-to-average power ratio is
limited. This result indicates that operating at very low $\tsnr$ is
power inefficient,
and the minimum bit energy of $M$-ary OOFSK signaling is achieved
at a nonzero spectral efficiency. Proposition
\ref{prop:finitePARslope} stems from the non-concavity of the
capacity-cost function under peak-to-average constraints (see
\cite{Verdu}). The minimum energy per bit must be computed
numerically.

\begin{figure}
\begin{center}
\includegraphics[width = 0.6\textwidth]{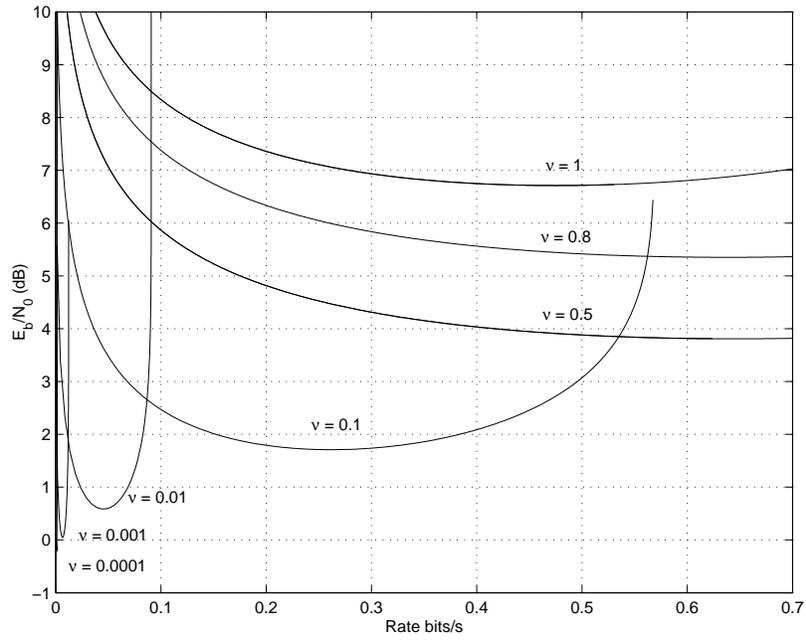}
\caption{$\frac{E_b}{N_0} \, (\textrm{\footnotesize{dB}})$ vs.
Rate bits/s for the unfaded Gaussian channel. $M=2$.}
\label{fig:peakyfskM2gauss}
\end{center}
\end{figure}

\begin{figure}
\begin{center}
\includegraphics[width = 0.6\textwidth]{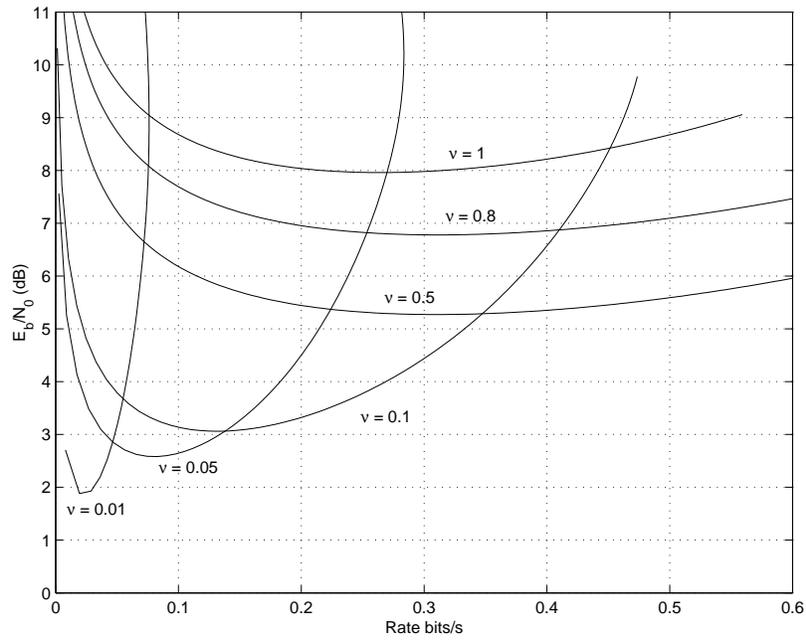}
\caption{$\frac{E_b}{N_0} \, (\textrm{\footnotesize{dB}})$ vs.
Rate bits/s for the unknown Rician channel with $\K = 0.5$.
$M=2$.} \label{fig:peakyfskM2alpha1sigma1}
\end{center}
\end{figure}

Figure \ref{fig:peakyfskM2gauss} plots bit energy curves as a
function of rate in bits/s achieved in the unfaded Gaussian channel
by 2-OOFSK signaling for different values of fixed duty factor
$\nu$. Notice that for all cases minimum bit energy values are
obtained at a nonzero rate and as the duty factor is decreased, the
required minimum bit energy is also decreased. With $\nu = 0.0001$,
the minimum bit energy is about $-0.2$ dB. Note that this is a
significant improvement over the case $\nu =1$ where the minimum bit
energy is about $6.7$ dB. However, this gain is obtained at the cost
of a considerable increase in the peak-to-average ratio. Fig.
\ref{fig:peakyfskM2alpha1sigma1} plots the bit energy curves in the
unknown Rician channel with Rician factor $\K = 0.5$.


\subsection{Limited Peak Power} \label{subsec:limitedpeak}

In this section, we consider the case where the peak level of the
transmitted signal is limited while there is no constraint on the
peak-to-average power ratio. Hence we fix the peak level to the
maximum allowed level, $A = \frac{P}{\nu}$. Therefore as $P \to 0$,
the duty factor also has to vanish and hence the peak-to-average
ratio increases without bound. In this case, the minimum bit energy
is achieved at zero spectral efficiency and the wideband slope
provides a good characterization of the bandwith/power tradeoff at
low spectral efficiency values.

\begin{cor:fixedPeakLimit}
Assume that the transmitter is limited in peak power,
$\frac{P}{\nu} \le A$, and the symbol duration $T$ is fixed. Then
the capacity achieved by $M$-ary OOFSK signaling, with fixed peak
power $A$, is a concave function of $P$. For the perfect receiver
side information case the minimum received bit energy and the
wideband slope are
\begin{gather}
\frac{E_b^r}{N_0}_{\min} = \frac{\log_e2}{\frac{E_{|h|}E_R\{\log
I_0(2\sqrt{\eta |h|^2 R })\}}{\eta (\gamma^2 + |d|^2)} - 1} \\
\intertext{and} S_0 = \frac{2 \left(E_h E_R \left\{ \log
I_0\left(2 \sqrt{\eta|h|^2R}\right) \right\} -  \eta (\gamma^2 +
|d|^2 )\right)^2} {E_h\{I_0(2 \eta |h|^2)\} - 1},
\end{gather}
respectively, where $R$ is a noncentral chi-square random variable
with
\begin{gather*}
p_{R} = e^{-R-\eta |h|^2} I_0(2\sqrt{\eta |h|^2 R})
\end{gather*}
and $\eta = A \frac{T}{N_0}$ is the normalized peak power. For the
imperfect receiver side information case the minimum received bit
energy and the wideband slope are
\begin{gather}
\frac{E_b^r}{N_0}_{\min} = \frac{\log_e2}{1 - \frac{1}{\gamma^2 +
|d|^2} \left( \frac{2|d|^2}{\eta \gamma^2 + 1} + \frac{\log(\eta
\gamma^2 + 1)}{\eta} - \frac{E\left\{\log I_0 \left( \frac{2
\sqrt{\eta |d|^2 R}}{\eta \gamma^2 +1} \right)\right\}}{\eta}
\right) } \label{eq:firstderivpeak}
\\
\intertext{and} S_0 = \left\{
\begin{array}{ll}
\frac{2\left( \eta(\gamma^2 + |d|^2) -
\frac{2\eta|d|^2}{\eta\gamma^2 + 1} \, - \, \log(\eta \gamma^2 +
1) + E\left\{\log I_0 \left( \frac{2 \sqrt{\eta |d|^2 R}}{\eta
\gamma^2 +1} \right)\right\} \right)^2}
{\frac{1}{1-\eta^2\gamma^4} \exp\left( \frac{2\eta^2 \gamma^2
|d|^2}{1 - \eta^2 \gamma^4} \right) I_0 \left( \frac{2\eta
|d|^2}{1-\eta^2\gamma^4} \right) - 1} & \eta \gamma^2 < 1
\\ 0 & \eta \gamma^2 \ge 1,
\end{array} \right. \label{eq:secondderivpeak}
\end{gather}
respectively, where $R$ is a noncentral chi-square random variable
with
\begin{gather*}
p_{R} = \frac{1}{\eta \gamma^2 + 1} \exp\left( -\frac{R + \eta
|d|^2 }{\eta \gamma^2 + 1}\right) I_0 \left( \frac{2 \sqrt{\eta
|d|^2 R} }{\eta \gamma^2 + 1} \right).
\end{gather*}
\end{cor:fixedPeakLimit}
\noindent \textbf{Proof}: Since perfect and imperfect receiver
side information cases are similar, for brevity we prove only the
latter case. When we fix the peak power $A = \frac{P}{v}$, we have
$v = \frac{\tsnr}{\eta}$ and the capacity becomes
\begin{align}
C_M^{ip}(\ssnr) = \left(1-\frac{\ssnr}{\eta}\right) \int p_{\R|X =
0} \log \frac{p_{\R|X=0}}{p_{\R}} \, \ud \R + \frac{\ssnr}{\eta}
\int p_{\R|X = 1} \log \frac{p_{\R|X=1}}{p_{\R}} \, \ud \R.
\nonumber
\end{align}
In the above capacity expression $p_{\R} = \left(1 -
\frac{\ssnr}{\eta} \right ) p_{\R | X = 0} + \frac{\ssnr}{M\eta}
\sum_{i = 1}^M p_{\R | X = i}$ where $p_{\R| X = 0}$ and $p_{\R| X
= i} $ for $1 \le i \le M$ do not depend on {\ssnr} because the
ratio $\frac{\ssnr}{\nu} = \eta$ is a constant. Concavity of the
capacity follows from the concavity of $-x \log x$ and the fact
that $p_{\R}$ is a linear function of $\ssnr$. Since the capacity
curve is concave, the minimum received bit energy is achieved at
zero spectral efficiency, $\frac{E_b^r}{N_0}_{\min} =
\frac{E\{|h|^2\} \log_e2}{\dot{C}(0)}$. The wideband slope is
given by (\ref{eq:widebandslope}), and depends on both the first
and second derivatives of the capacity.  Hence the expressions in
(\ref{eq:firstderivpeak}) and (\ref{eq:secondderivpeak}) are
easily obtained by evaluating
\begin{gather}
\dot{C}_M^{ip}(0) = \gamma^2 + |d|^2 - \frac{2 |d|^2}{\eta \gamma^2
+ 1} - \frac{\log(\eta \gamma^2 +1)}{\eta} + \frac{E\left\{ \log I_0
\left( \frac{2 \sqrt{\eta |d|^2 R}}{\eta \gamma^2 + 1} \right)
\right\}}{\eta} \nonumber \intertext{and} \ddot{C}_M^{ip}(0) =
\left\{
\begin{array}{ll}
\frac{1}{\eta^2 M} \left( 1 - \frac{1}{1 - \eta^2 \gamma^4}
\exp\left( \frac{2 \eta^2 \gamma^2 |d|^2}{1 - \eta^2 \gamma^4}
\right) I_0 \left( \frac{2\eta |d|^2}{1-\eta^2\gamma^4} \right)
\right) & \eta \gamma^2 < 1  \\ -\infty & \eta \gamma^2 \ge 1.
\end{array} \right. \label{eq:secondderivpeakintheproof}
\end{gather}
Similarly, for the perfect receiver side information case, we note
that
\begin{gather*}
\dot{C}_M^{p}(0) = \frac{E_{|h|}E_R \{ \log I_0 (2\sqrt{\eta |h|^2 R
})\}}{\eta} - (\gamma^2 +|d|^2) \intertext{and} \ddot{C}_M^p(0) =
\frac{1 - E_{|h|} \{ I_0(2\eta |h|^2) \}}{\eta^2 M}.
\end{gather*}
\hfill $\square$

In contrast to the limited PAR case, the minimum bit energy is
achieved at zero spectral efficiency, and hence the power efficiency
of the system improves if one operates at smaller {\tsnr} and
vanishing duty factor. Note in this case that, although the average
power $P$ is decreasing, the energy of FSK signals,
$\frac{PT}{\nu}$, is kept fixed, and the average power constraint is
satisfied by sending these signals less frequently. In the
imperfectly known channel, this type of peakedness introduced in
time proves useful in avoiding adverse channel conditions. On the
other hand, in the PAR limited case, the decreasing average power
constraint is satisfied by decreasing the energy of FSK signals.
Note that in the above result, for both perfect and imperfect side
information cases, the minimum bit energy and the wideband slope do
not depend on $M$. Therefore On/Off signaling with vanishing duty
cycle is optimally power-efficient at very low spectral efficiency
values, and there is no need for frequency modulation. Further note
that in the imperfect receiver side information case, if $\eta
\gamma^2 \ge 1$, then $S_0 = 0$, and hence approaching the minimum
bit energy is extremely slow. If we relax the peak power limitation
and let $\eta \uparrow \infty$, then it is easily seen that even in
the imperfect receiver side information case,
$\frac{E_b^r}{N_0}_{\min} \to \log_e2 = -1.59$ dB. Indeed,
\cite{Verdu} shows in a more general setting that flash signaling
with increasingly high peak power is required to achieve the minimum
bit energy of $-1.59$ dB if the fading is not perfectly known at the
receiver.

\begin{figure}
\begin{center}
\includegraphics[width = 0.6\textwidth]{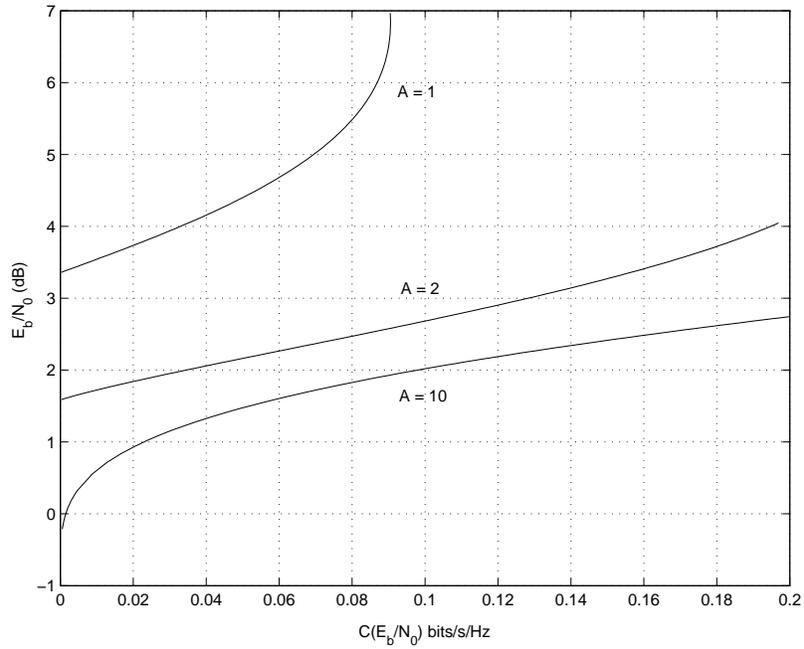}
\caption{$\frac{E_b}{N_0} \, (\textrm{\footnotesize{dB}})$ vs. vs.
Spectral Efficiency $C(\frac{E_b}{N_0})$ bits/s/Hz for the unfaded
Gaussian channel. $M=2$.} \label{fig:unboundedPARM2gauss}
\end{center}
\end{figure}

Fig. \ref{fig:unboundedPARM2gauss} plots the bit energy curves
achieved by $2$-OOFSK signaling in the unfaded Gaussian channel for
different peak power values $A$. Notice that for all cases the
minimum bit energy is achieved in the limit as the spectral
efficiency goes to zero and this energy monotonically decreases to
$-1.59$ dB as $A \to \infty$.


\section{Capacity of $M$-ary OOFPSK Signaling}\label{sec:oofskwithphase}

In this section, we consider joint frequency and phase modulation to
improve the power efficiency of communication with OOFSK signaling.
Combining phase and frequency modulation techniques has been
proposed in the literature (see e.g., \cite{Padovani},
\cite{Ghareeb}, \cite{Khalona}, and \cite{Hung}). As we have seen in
the previous section, if the receiver employs energy detection and
the peak-to-average power ratio is limited, then operating at very
low $\tsnr$ is extremely power inefficient. The peak-to-average
power ratio constraint puts a restriction on the energy
concentration in a fraction of time. Hence, for low average power
values, the power of FSK signals is also low, and depending solely
on energy detection leads to severe degradation in the performance.
On the other hand, if the receiver can track phase shifts in the
channel or if the received signal has a specular component as in the
Rician channel, then the performance is improved at low spectral
efficiency values if information is conveyed in not only the
amplitude but also the phase of each orthogonal frequency. Hence we
propose employing phase modulation in OOFSK signaling. Therefore, in
this section, we assume that the phase $\theta_i$ of the FSK signal
\begin{gather}
s_{i, \theta_i}(t) = \sqrt{\frac{P}{\nu}} \, e^{j(w_it +
\theta_i)} \quad 0\le t \le T
\end{gather}
is a random variable carrying information. Henceforth this new
signaling scheme is referred to as OOFPSK signaling. The channel
input can now be represented by the pair $(X,\theta)$. If $X = i$
for $1 \le i \le M $, and $\theta = \theta_i$, the transmitter sends
the sine wave $s_{i,\theta_i}(t)$, while no transmission is denoted
by $X = 0$, and hence $s_0(t) = 0$. As another difference from
Section \ref{sec:oofsk} ,the decoder directly uses the matched
filtered output vector $\Y = (Y_1, \ldots, Y_M)$ instead of the
energy measurements in each frequency component.

\subsection{Perfect  Receiver Side Information}

We first consider the case where the receiver has perfect knowledge
of the instantaneous realization of fading coefficients $\{h_k\}$,
and obtain the capacity results both for fixed $M$ and as $M$ goes
to infinity.

\begin{prop:coherentoofskcapwithphase} \label{prop:coherentoofskcapwithphase}
Consider the fading channel model (\ref{eq:oofskmodel}) and assume
that the receiver perfectly knows the instantaneous values of the
fading, $h_k$, $k = 1,2,\ldots$ while the transmitter has no
fading side information. Then the capacity of $M$-ary orthogonal
OOFPSK signaling with a fixed duty factor $\nu \le 1$ is
\begin{gather}
C_{M}^p(\ssnr) = -M - E_{|h|}\left\{ (1-\nu)\int p_{\R| X = 0}
\log p_{\R | \, |h| } \, \ud \R + \nu \int p_{\R |X = 1,|h|} \log
p_{\R | \, |h|} \, \ud \R\right\}
\label{eq:coherentoofskcapwithphase}
\end{gather}
where $p_{\R | \, |h|}$, $p_{\R|X = 0}$, $p_{\R|X = i, |h|}$ and
$f(R_i,|h|, \ssnr)$ for $1 \le i \le M$ are defined in
(\ref{eq:p(R)}), (\ref{eq:p(R|X=0)}), (\ref{eq:p(R|X = i)}) and
(\ref{eq:fRi}) respectively.
\end{prop:coherentoofskcapwithphase}
\noindent \textbf{Proof}: See Appendix
\ref{appendix:coherentoofskcapwithphase}.

\begin{cor:coherentcapMinfwithphase}
The capacity expression (\ref{eq:coherentoofskcapwithphase}) of
$M$-ary OOFPSK signaling in the limit as $M \uparrow \infty$
becomes
\begin{align}
C_{\infty}^p (\ssnr) &= D(P_{y|\tilde{x},h} \big \|
P_{y|\tilde{x}=0, h} \big | F_{\tilde{x}}F_h ) \nonumber \\ &=
E\{|h|^2\} \, \ssnr \nonumber
\\
&= (\gamma^2 + |d|^2) \ssnr,
\end{align}
where $y = h\tilde{x} + n$, $\tilde{x}$ is a two-mass-point
discrete random variable with mass-point locations and
probabilities given in (\ref{eq:tildex}), and $n$ is zero mean
circularly symmetric Gaussian random variable with $E\{|n|^2\} =
1$.
\end{cor:coherentcapMinfwithphase}

Note that $\frac{1}{T}\, C_{\infty}^p (\ssnr) = (\gamma^2 + |d|^2)
\frac{P}{N_0}$ nats/s  is equal to the infinite bandwidth capacity
of the unfaded Gaussian channel with the same received power.
Hence, in the perfect side information case ordinary FPSK
signaling with duty factor $\nu =1$ is enough to achieve this
capacity.


\subsection{Imperfect Receiver Side Information}

Similarly as in Section \ref{subsec:imperfect}, we now assume that
neither the receiver nor the transmitter has any fading side
information and consider a more special fading process: memoryless
Rician fading where each of the i.i.d. $h_k$'s is a proper complex
Gaussian random variable with $E\{h_k\} = d$ and $\text{var}(h_k)
= \gamma^2$. The capacity of OOFPSK signaling is given by the
following result.

\begin{prop:oofskcapwithphase}
Consider the fading channel (\ref{eq:oofskmodel}) and assume that
the fading process $\{h_k\}$ is a sequence of i.i.d. proper
complex Gaussian random variables with $E\{h_k\} = d$ and
$\text{var}(h_k) = \gamma^2$ which are not known at either the
receiver or the transmitter. Then the capacity of $M$-ary
orthogonal OOFPSK signaling with a duty factor $\nu \le 1$ is
given by
\begin{align}
C_{M}^{ip}(\ssnr) = &-M - \nu \log(\gamma^2 \ssnr / \nu + 1) -
(1-\nu) \int p_{\R|X = 0} \log p_{\R} \, \ud \R \nonumber \\ &-\nu
\int p_{\R|X = 1} \log p_{\R} \, \ud \R
\label{eq:oofskcapwithphase}
\end{align}
where $p_{\R}$, $p_{\R|X = 0}$, $p_{\R|X = i}$ and $f(R_i,\ssnr)$
for $1 \le i \le M$ are defined in (\ref{eq:p(R)imperfect}),
(\ref{eq:p(R|X=0)imperfect}), (\ref{eq:p(R|X=i)imperfect}) and
(\ref{eq:fRiimperfect}) respectively.
\end{prop:oofskcapwithphase}
\noindent\textbf{Proof}: The proof is almost identical to that of
Proposition \ref{prop:coherentoofskcapwithphase}. Due to the
symmetry of the channel, capacity is achieved by equiprobable FSK
signals with uniform phases. Note that in this case,
\begin{align*}
C_{M}^{ip}(\ssnr) = &(1-\nu) \int p_{\Y|X = 0,\theta} \log
\frac{p_{\Y|X=0,\theta}}{p_{\Y}} \, \ud \Y \, \frac{1}{2 \pi} \,
\ud \theta \\ &+ \nu \int p_{\Y|X = 1,\theta} \log
\frac{p_{\Y|X=1,\theta}}{p_{\Y}} \, \ud \Y \, \frac{1}{2 \pi} \,
\ud \theta
\end{align*}
where
\begin{align*}
p_{\Y | X = i, \theta_i} = \left\{
\begin{array}{ll}
\frac{1}{\pi^{M-1}} e^{-\sum_{j \neq i} |Y_j|^2}
\frac{1}{\pi(\gamma^2 \alpha^2 + 1 )} e^{-\frac{|Y_i - \alpha d
e^{j\theta_i} |^2}{\gamma^2 \alpha^2 +1 }} & 1 \le i \le M \\
\frac{1}{\pi^M} e^{-\sum_{j = 1}^M |Y_j|^2} & i = 0.
\end{array} \right.
\end{align*}
The capacity expression in (\ref{eq:oofskcapwithphase}) is then
obtained by first integrating with respect to $\theta$ and then
making a change of variables, $R_j = |Y_j|^2$. \hfill $\square$

\begin{cor:capMinfwithphase}
The capacity expression (\ref{eq:oofskcapwithphase}) of $M$-ary
OOFPSK signaling in the limit as $M \uparrow \infty$ becomes
\begin{align}
C_{\infty}^{ip}(\ssnr) &= D(P_{y|\tilde{x}} \big \| P_{y|\tilde{x} =
0} \big | F_{\tilde{x}}) \nonumber
\\
&= (\gamma^2 + |d|^2) \, \ssnr - \nu \log \left( \gamma^2
\frac{\ssnr}{\nu} + 1 \right), \label{eq:Minfnoncohwithphase}
\end{align}
where $y = h\tilde{x} + n$, $h$ is a proper Gaussian random
variable with $E\{h\} = d$ and $\text{var}(h) = \gamma^2$,
$\tilde{x}$ is a two-mass-point discrete random variable with
mass-point locations and probabilities given in (\ref{eq:tildex}),
and $n$ is a zero mean circularly symmetric complex Gaussian
random variable with $E\{|n|^2\} = 1$.
\end{cor:capMinfwithphase}

\noindent Similarly as before, the remarks below are given for the
asymptotic case in which $M \to \infty$.

\begin{rem:cohnoncohlosswithphase}
\emph{Assume that in the case of perfect receiver side
information, $\{h_k\}$ is a sequence of i.i.d. proper complex
Gaussian random variables. Then the asymptotic loss in capacity
incurred by not knowing the fading is
\begin{align}
C_{\infty}^p(\ssnr) - C_{\infty}^{ip}(\ssnr) &= D(p_{y
|\,\tilde{x}, h} \, \big\| \, p_{y| \, \tilde{x} = 0, h } \big| \,
F_{h} F_{\tilde{x}}) -  D(p_{y|\tilde{x}} \, \big\| \, p_{y |
\tilde{x}=0 } \big| F_{\tilde{x}}) \nonumber
\\
& = I(h ; y \, \big| \, \tilde{x}).
\end{align}}
\end{rem:cohnoncohlosswithphase}

\begin{rem:Minfasymptoticswithphase}
\emph{Consider the case of imperfect receiver side information.
For unit duty factor $\nu =1$, the capacity expression
(\ref{eq:Minfnoncohwithphase}) is a special case of the result by
Viterbi \cite{Viterbi}. From (\ref{eq:Minfnoncohwithphase}) we can
also see that for fixed symbol interval $T$,
\begin{gather}\label{eq:Minfnoncohvwithphase}
\lim_{\nu \downarrow 0} \frac{1}{T} \, C_{\infty}^{ip}(\ssnr) =
\frac{1}{T}(\gamma^2 + |d|^2) \ssnr = (\gamma^2 + |d|^2)
\frac{P}{N_0} \text{ nats/s},
\end{gather}
and for fixed duty factor $\nu$,
\begin{gather}\label{eq:MinfnoncohTwithphase}
\lim_{T \uparrow \infty} \frac{1}{T} \, C_{\infty}^{ip}(\ssnr) =
(\gamma^2 + |d|^2) \frac{P}{N_0} \text{ nats/s}.
\end{gather}
Note that right-hand sides of (\ref{eq:Minfnoncohvwithphase}) and
(\ref{eq:MinfnoncohTwithphase}) are equal to the infinite
bandwidth capacity of the unfaded Gaussian channel with the same
received signal power.}
\end{rem:Minfasymptoticswithphase}


\subsection{Limited Peak-to-Average Power Ratio}

As in Section \ref{subsec:limitedPAR}, we first consider the case
where the tranmitter peak-to-average power ratio is limited and
hence the duty factor $\nu$ is kept fixed while the average power
varies. The power efficiency in the low-power regime is
characterized by the following result.

\begin{cor:finiteParSlopewithphase} \label{cor:finiteParSlopewithphase}
Assume that the transmitter is constrained to have limited peak to
average power ratio and the PAR of $M$-ary OOFPSK signaling,
$1/\nu$, is kept fixed at its maximum level. Then for the perfect
receiver side information case the minimum received bit energy  and
the wideband slope are
\begin{gather}
\frac{E_b^r}{N_0}_{\min} = \log_e2 \quad \text{and} \quad S_{0} =
\frac{2 \left( E\{|h|^2\}\right)^2}{E\{|h|^4\}} =
\frac{2}{\kappa(|h|)} \label{eq:slopeperfectwithphase}
\end{gather}
respectively, where $\kappa(|h|)$ is the kurtosis of the fading
magnitude. For the imperfect receiver side information case, the
received bit energy required at zero spectral efficiency and the
wideband slope are
\begin{gather}
\left. \frac{E_b^r}{N_0} \right|_{\C = 0} = \left( 1 +
\frac{1}{\K} \right) \log_e2 \quad \text{and} \quad S_0 = \frac{2
\K^2} {\left( 1 + \K\right)^2 - \frac{M}{\nu}}
\label{eq:slopeimperfectwithphase}
\end{gather}
respectively, where $\K = \frac{|d|^2}{\gamma^2}$ is the Rician
factor.
\end{cor:finiteParSlopewithphase}
\noindent \textbf{Proof}: For brevity, we show the result only for
the imperfect receiver side information case. Note that in the
capacity expression (\ref{eq:oofskcapwithphase}), the only term
that depends on {\tsnr} is $f(R_i,\tsnr)$. Using $$\lim_{x \to 0}
\frac{I_1(a\sqrt{x})}{\sqrt{x}} = \frac{a}{2}$$ and $$\lim_{x \to
0} \frac{I_0(a\sqrt{x})}{x} - \frac{2I_1(a\sqrt{x})}{ax^{3/2}} =
\frac{a^2}{8},$$ one can easily show that the first and second
derivatives with respect to $\tsnr$ of $f(R_i,\tsnr)$ at zero
$\tsnr$ are
\begin{gather*}
\dot{f}(R_i,0) = \frac{1}{\nu} (\gamma^2 + |d|^2)(-1 + R_i)
\intertext{and} \ddot{f}(R_i,0) = \frac{1}{\nu^2}(|d|^4 +
2\gamma^4 + 4\gamma^2 |d|^2)\left(1 - 2 R_i + \frac{R_i^2}{2}
\right),
\end{gather*}
respectively. Then, differentiating the capacity
(\ref{eq:oofskcapwithphase}) with respect to $\tsnr$ we have
\begin{gather} \label{eq:oofskPARderiv}
\dot{C}_M^{ip}(0) = |d|^2 \quad \text{and} \quad \ddot{C}_M^{ip}(0)
= -\frac{(\gamma^2 + |d|^2)^2}{M} + \frac{\gamma^4}{\nu}.
\end{gather}
The received bit energy required at zero spectral efficiency is
obtained from the formula $$\left. \frac{E_b^r}{N_0} \right|_{\C =
0} = \frac{(\gamma^2 + |d|^2) \log_e2}{\dot{C}(0)}$$ and the
wideband slope is found by inserting the derivative expressions in
(\ref{eq:oofskPARderiv}) into (\ref{eq:widebandslope}). Similarly,
for the perfect receiver side information case, we have
\begin{gather*}
\dot{C}_M^p(0) = E\{|h|^2\} = (\gamma^2 + |d|^2) \quad \text{and}
\quad \ddot{C}_M^p(0) = - \frac{E\{|h|^4\}}{M}.
\end{gather*}
\hfill $\square$

\noindent Notice that in the perfect side information case, the
minimum bit energy is $-1.59$ dB and the wideband slope does not
depend on $M$ and $\nu$. In fact, Verd\'u has obtained the same
bit energy and wideband slope expression in \cite{Verdu} for
discrete-time fading channels when the receiver knows the fading
coefficients, and proved that QPSK modulation is optimally
efficient achieving these values. More interesting is the
imperfect receiver side information case, where the minimum bit
energy is not necessarily achieved at zero spectral efficiency.
Note that unlike the bit energy expression in
(\ref{eq:slopeimperfectwithphase}), the wideband slope is a
function of $M$ and $\nu$ and is negative if $\frac{M}{\nu} > (1 +
\K)^2$ in which case the minimum bit energy is achieved at a
nonzero spectral efficiency.

\begin{figure}
\begin{center}
\includegraphics[width = 0.6\textwidth]{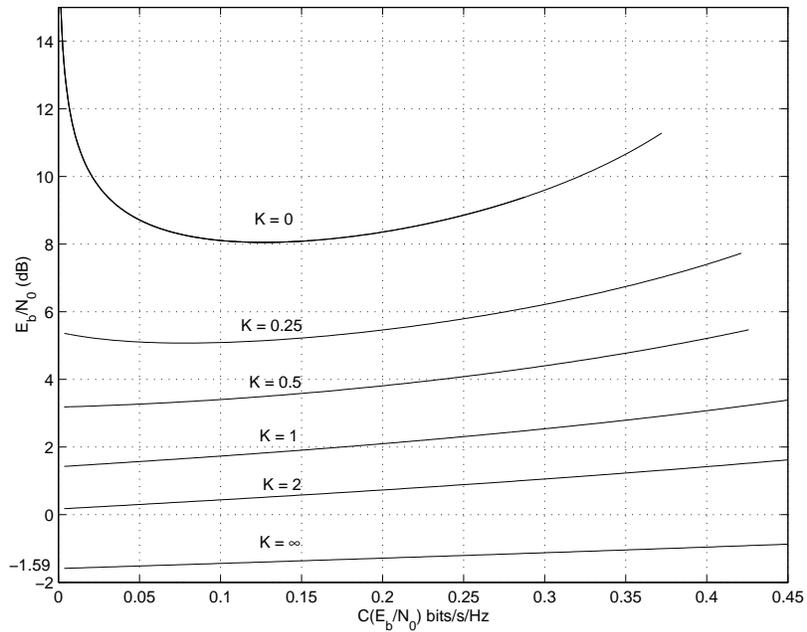}
\caption{$\frac{E_b}{N_0} \, (\textrm{\footnotesize{dB}})$ vs.
Spectral Efficiency $C(\frac{E_b}{N_0})$ bits/s/Hz for the unknown
Rayleigh channel ($\K = 0$), unknown Rician channels ($\K =
0.25,0.5,1,2$) and the unfaded Gaussian channel ($\K = \infty$) when
$M=2$ and $\nu = 1$.} \label{fig:fskpsk_M2}
\end{center}
\end{figure}

\begin{figure}
\begin{center}
\includegraphics[width = 0.6\textwidth]{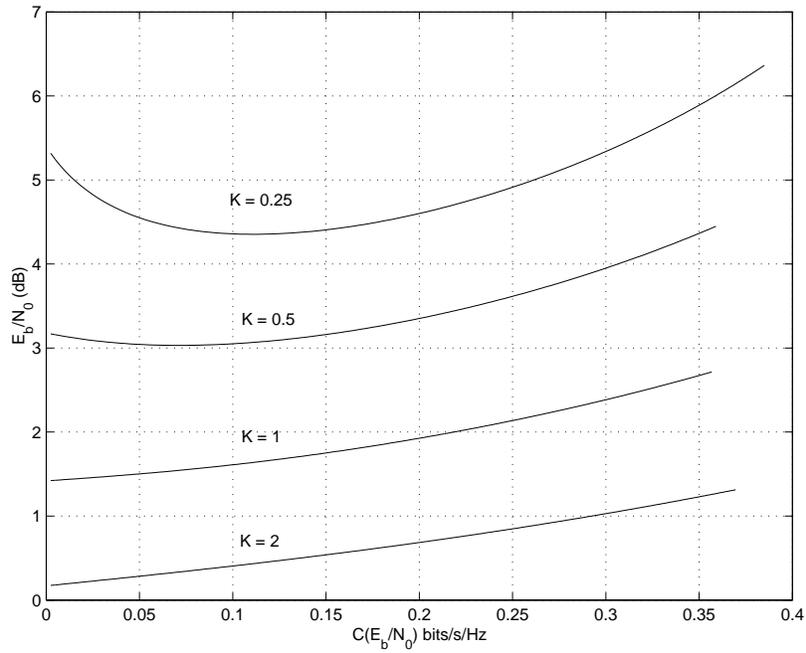}
\caption{$\frac{E_b}{N_0} \, (\textrm{\footnotesize{dB}})$ vs.
Spectral Efficiency $C(\frac{E_b}{N_0})$ bits/s/Hz for unknown
Rician channels ($\K = 0.25,0.5,1,2$) when $M=3$ and $\nu = 1$.}
\label{fig:fskpsk_M3}
\end{center}
\end{figure}

\begin{figure}
\begin{center}
\includegraphics[width = 0.6\textwidth]{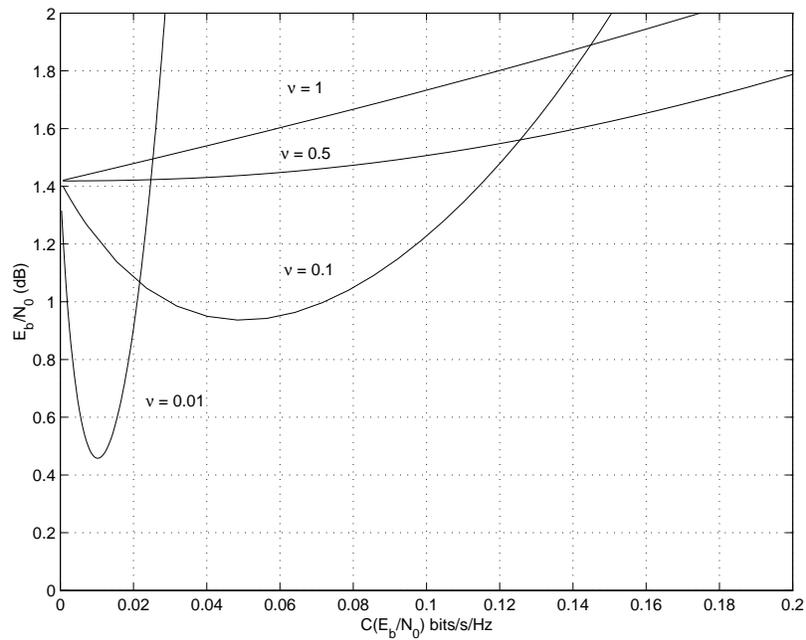}
\caption{$\frac{E_b}{N_0} \, (\textrm{\footnotesize{dB}})$ vs.
Spectral Efficiency $C(\frac{E_b}{N_0})$ bits/s/Hz for the unknown
Rician channel with $\K = 1$ for $\nu = 1, 0.5, 0.1, 0.01$ when
$M=2$.} \label{fig:fskpsk_M2K1various_v}
\end{center}
\end{figure}

Figure \ref{fig:fskpsk_M2} plots the bit energy curves as a
function of spectral efficiency in bits/s/Hz for 2-FPSK signaling
($\nu = 1$). Note that for $\K = 0.25$, the wideband slope is
negative, and hence the minimum bit energy is achieved at a
nonzero spectral efficiency. On the other hand for $K = 0.5,1,2$,
the wideband slope is positive, and hence higher power efficiency
is achieved as one operates at lower spectral efficiency. Similar
observations are noted from Fig. \ref{fig:fskpsk_M3} where bit
energy curves are plotted for 3-FPSK signaling. Fig.
\ref{fig:fskpsk_M2K1various_v} plots the bit energy curves for
2-OOFPSK signaling with different duty cycle parameters over the
unknown Rician channel with $\K = 1$. We observe that the required
minimum bit energy is decreasing with decreasing duty cycle. For
instance, when $\nu = 0.01$, the minimum bit energy of $\sim
0.46$dB is achieved at the cost of a peak-to-average ratio of 100.
Note also that since the received bit energy at zero spectral
efficiency (\ref{eq:slopeimperfectwithphase}) depends only on the
Rician factor $\K$, all the curves in Fig.
\ref{fig:fskpsk_M2K1various_v} meet at the same point on the
$y$-axis.


\subsection{Limited Peak Power}

Here we assume that the transmitter is limited in its peak power
while there is no bound on the peak-to-average power ratio. We
consider the power efficiency of $M$-ary OOFPSK signaling when the
peak power is kept fixed at the maximum allowed level, $A =
\frac{P}{\nu}$. Note that as the average power $P \to 0$, the duty
factor $\nu$ also must vanish, thereby increasing the
peak-to-average power ratio without bound. For this case, we have
the following result.

\begin{cor:fixedPeakLimitwithphase}
Assume that the transmitter is limited in peak power,
$\frac{P}{\nu} \le A$, and the symbol duration $T$ is fixed. Then
the capacity achieved by $M$-ary OOFPSK signaling with fixed peak
power $A$ is a concave function of the $\tsnr$. For the case of
perfect receiver side information, the minimum received bit energy
and the wideband slope are
\begin{gather}
\frac{E_b^r}{N_0}_{\min} = \log_e2 \,\, \text{ and } \,\, S_{0} =
\frac{2 {\eta^2} \left( E\{|h|^2\} \right)^2} {E\{I_0(2\eta
|h|^2)\}- 1}, \label{eq:peakslopecohwithphase}
\end{gather}
respectively, where 
$\eta = A \frac{T}{N_0}$ is the normalized peak power. For the
case of imperfect receiver side information, the minimum received
bit energy and the wideband slope are
\begin{gather}
\frac{E_b^r}{N_0}_{\min} = \frac{\log_e2}{1 - \frac{\log(\gamma^2
\eta + 1 )}{(\gamma^2 + |d|^2)\eta}} \quad \text{and} \quad S_0 =
\left\{
\begin{array}{ll}
\frac{2\left( \eta(\gamma^2 + |d|^2) - \log(\eta \gamma^2 + 1)
\right)^2} {\frac{1}{1-\eta^2\gamma^4} \exp\left( \frac{2\eta^2
\gamma^2 |d|^2}{1 - \eta^2 \gamma^4} \right) I_0 \left(
\frac{2\eta |d|^2}{1-\eta^2\gamma^4} \right) - 1} & \eta \gamma^2
< 1
\\ 0 & \eta \gamma^2 \ge 1 \label{eq:peakslopenoncohwithphase}
\end{array} \right.
\end{gather}
respectively. 
\end{cor:fixedPeakLimitwithphase}
\noindent \textbf{Proof}: As before, we consider only the
imperfect receiver side information case. When we fix the peak
power $A = \frac{P}{v}$, we have $v = \frac{\tsnr}{\eta}$ and the
capacity becomes
\begin{align*}
C_{M}^{ip}(\ssnr) = &- M - \frac{\ssnr}{\eta} \log(\gamma^2 \eta +
1) - \left(1 - \frac{\ssnr}{\eta} \right) \int p_{\R|X = 0} \log
p_{\R} \, \ud \R \\ &- \frac{\ssnr}{\eta} \int p_{\R|X = 1} \log
p_{\R} \, \ud \R.
\end{align*}
In the above capacity expression $$p_{\R} = \left(1 -
\frac{\ssnr}{\eta} \right ) p_{\R | X = 0} + \frac{\ssnr}{M\eta}
\sum_{i = 1}^M p_{\R | X = i}$$ where $p_{\R| X = 0}$ and $p_{\R| X
= i} $ for $1 \le i \le M$ do not depend on {\ssnr} because the
ratio $\frac{\ssnr}{\nu} = \eta$ is a constant. Concavity of the
capacity follows from the concavity of $-x \log x$ and the fact that
$p_{\R}$ is a linear function of $\ssnr$. Due to concavity of the
capacity curve, the minimum bit energy is achieved at zero spectral
efficiency.  Differentiating the capacity with respect to $\ssnr$,
we get
\begin{gather*}
\dot{C}_M^{ip}(0) = \gamma^2 + |d|^2 - \frac{\log(\gamma^2 \eta +
1)}{\eta},
\end{gather*}
and $\ddot{C}_M^{ip}(0)$ having the same expression as in
(\ref{eq:secondderivpeakintheproof}). Then,
(\ref{eq:peakslopenoncohwithphase}) is easily obtained using the
aforementioned formulas for the minimum bit energy and the wideband
slope. Similarly, we note for the perfect side information case that
\begin{gather*}
\dot{C}_M^p(0) = E\{|h|^2\} = \gamma^2 + |d|^2 \quad \text{and}
\quad \ddot{C}_M^p(0) = \frac{1 - E\{I_0(2\eta |h|^2)\}}{\eta^2 M}.
\end{gather*}
\hfill $\square$

Note that the results in (\ref{eq:peakslopecohwithphase}) and
(\ref{eq:peakslopenoncohwithphase}) do not depend on $M$, and
hence they can be achieved by pure On/Off keying. Further note
that $\frac{I_0(2\eta |h|^2) - 1}{\eta^2} > |h|^4$ for $\eta
> 0$. Therefore, when the fading is perfectly known, the strategy
of fixing the peak power and letting $\nu \downarrow 0$ results in
a wideband slope smaller than that of fixed duty factor and hence
should not be preferred. In the imperfect receiver side
information case, if the peak power limitation is relaxed, i.e.,
$\eta \uparrow \infty$, the minimum bit energy approaches $-1.59$
dB.

\begin{figure}
\begin{center}
\includegraphics[width = 0.6\textwidth]{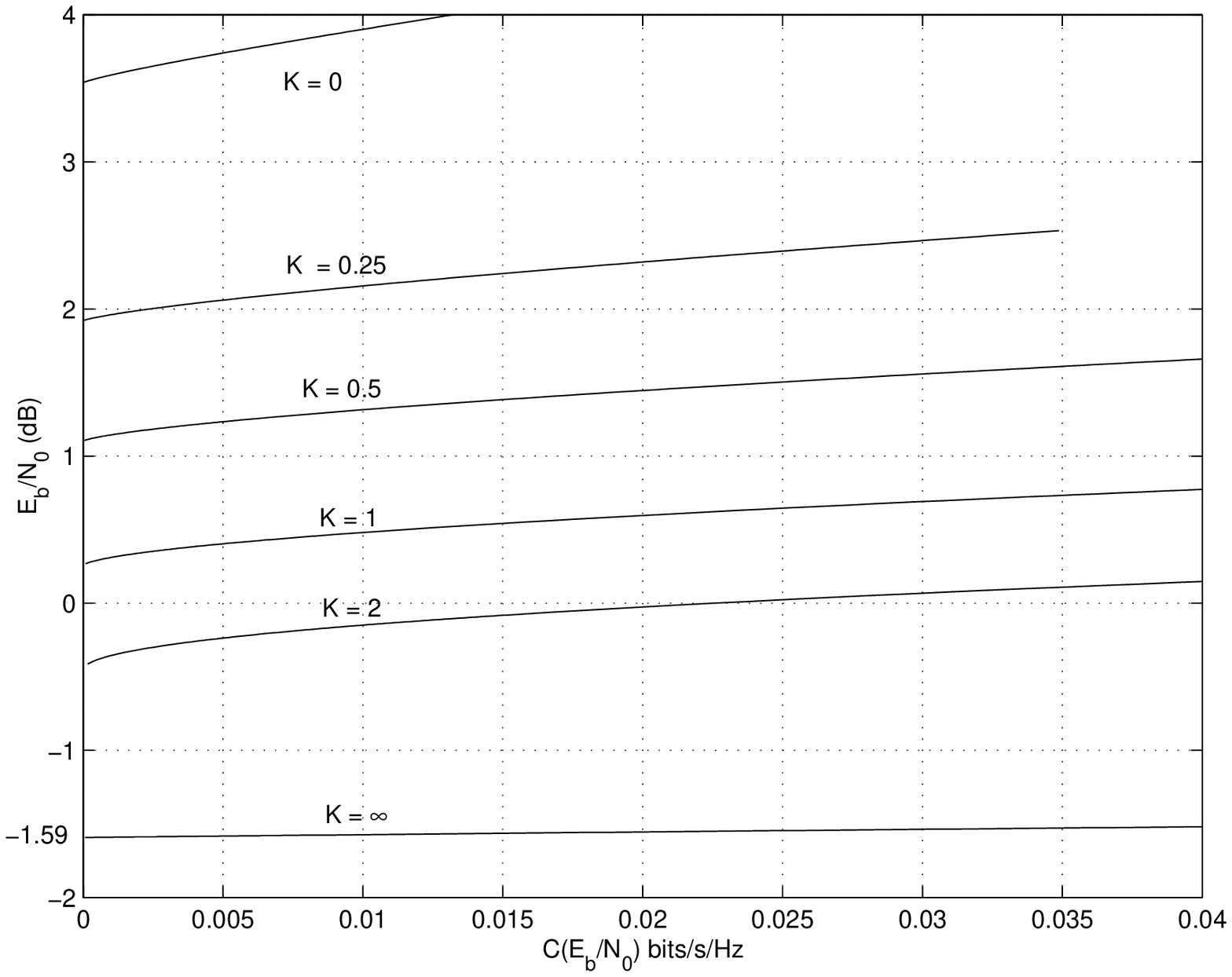}
\caption{$\frac{E_b}{N_0} \, (\textrm{\footnotesize{dB}})$ vs.
Spectral Efficiency $C(\frac{E_b}{N_0})$ bits/s/Hz for the unknown
Rayleigh channel ($\K = 0$), unknown Rician channels ($\K =
0.25,0.5,1,2$), and the unfaded Gaussian channel ($\K = \infty$)
when $M=2$ and fixed peak limit $\eta = 1$.}
\label{fig:fskpsk_M2fixedpeaketa_1}
\end{center}
\end{figure}
\begin{figure}
\begin{center}
\includegraphics[width = 0.6\textwidth]{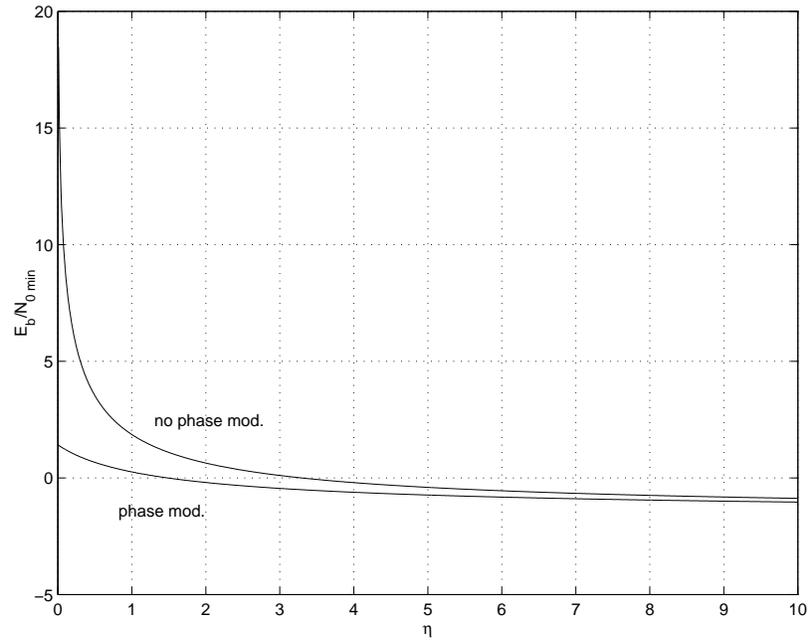}
\caption{$\frac{E_b}{N_0}_{\min}$ vs. normalized peak power limit
$\eta$ in the unknown Rician channel with $\K = 1$.}
\label{fig:EbNominK1}
\end{center}
\end{figure}
\begin{figure}
\begin{center}
\includegraphics[width = 0.6\textwidth]{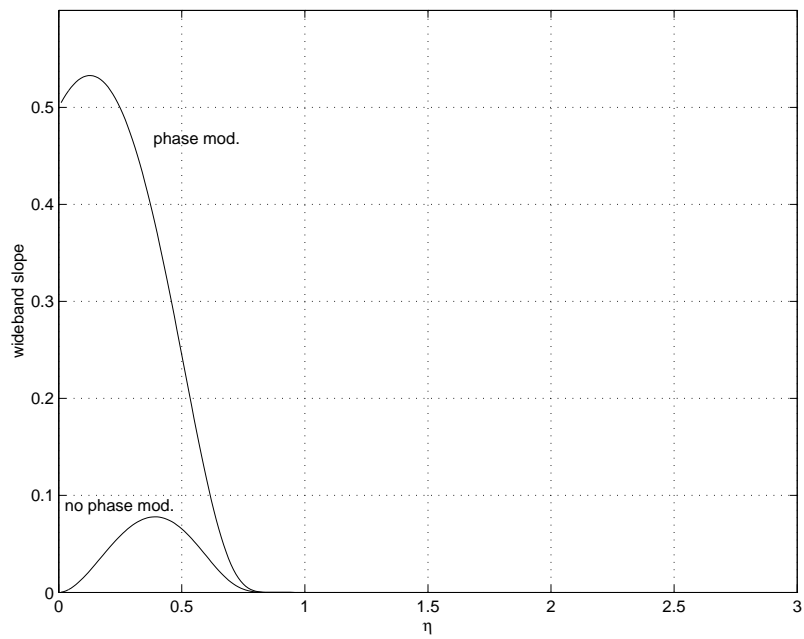}
\caption{Wideband Slope $S_0$ vs. normalized peak power limit
$\eta$ in the unknown Rician channel with $\K = 1$.}
\label{fig:S0K1}
\end{center}
\end{figure}

Fig. \ref{fig:fskpsk_M2fixedpeaketa_1} plots the bit energy curves
as a function of spectral efficiency for the unknown Rayleigh
channel ($\K = 0$), unknown Rician channels ($\K = 0.25, 0.5, 1,
2$), and the unfaded Gaussian channel ($\K = \infty$) when the
normalized peak power limit is $\eta = 1$. We observe that for all
cases the required bit energy decreases with decreasing spectral
efficiency, and therefore the minimum bit energy is achieved at
zero spectral efficiency. Finally Figures \ref{fig:EbNominK1} and
\ref{fig:S0K1} plot the minimum bit energy and wideband slope
values, respectively, as functions of the normalized peak power
limit $\eta$ in the unknown Rician channel with $\K = 1$. The
curves are plotted for the case in which no phase modulation is
used and the receiver employs energy detection (Section
\ref{sec:oofsk}), and also for the scenario in which phase
modulation is employed.


\section{Conclusion} \label{sec:oofskconclusion}

We have considered transmission of information over wideband fading
channels using $M$-ary orthogonal On/Off FSK (OOFSK) signaling, in
which $M$-ary FSK signaling is overlaid on top of On/Off keying. We
have first assumed that the receiver uses energy detection for the
reception of OOFSK signals. We have obtained capacity expressions
when the receiver has perfect and imperfect fading side information
both for fixed $M$ and as $M$ goes to infinity. We have investigated
power efficiency when the transmitter is subject to a
peak-to-average power ratio (PAR) limitation or a peak power
limitation. It is shown that under a PAR limitation no matter how
large the transmitted energy per information bit is, reliable
communication is impossible for small enough spectral efficiency
even in the unfaded Gaussian channel, and hence it is extremely
power inefficient to operate in the very low $\tsnr$ regime. On the
other hand, if there is only a peak power limitation, we have
demonstrated that power efficiency improves as one operates with
smaller $\tsnr$ and vanishing duty factor. We note that in this case
On/Off keying (OOK) is an optimally efficient signaling in the low
power regime achieving the minimum bit energy and the wideband slope
in both perfect and imperfect channel side information cases, while
combined OOK and FSK signaling is required to improve energy
efficiency when a constraint is imposed on the PAR.

We have also considered joint frequency-phase modulation schemes
where the phase of the FSK signals are also used to convey
information. Similarly we have analyzed the capacity and power
efficiency of these schemes. Assuming perfect channel knowledge at
the receiver, we have obtained the minimum bit energy and wideband
slope expressions. In this case, it is shown that FSK signaling is
not required for optimum power efficiency in the low-power regime as
pure phase modulation in the PAR limited case and OOK in the peak
power limited case achieve both the minimum bit energy and the
optimal wideband slope.  For the case in which the receiver has
imperfect channel side information and the input is subject to PAR
constraints, we have shown that if $\frac{M}{\nu}
> (1+\K)^2$, then the wideband slope is negative, and hence the
minimum bit energy is achieved at a nonzero spectral efficiency,
$\C^* > 0$. It is concluded that, in these cases, operating in the
region, where $\C < \C^*$, should be avoided. We also note that in
general the combined OOK and FSK signaling performs better and
indeed if the number of orthogonal frequencies, i.e., $M$, is
increased then a smaller minimum bit energy value is achieved.
Furthermore, for the case in which only the peak power is limited
with no constraints on the peak-to-average ratio, we have
investigated the spectral-efficiency/bit-energy tradeoff in the
low-power regime by obtaining both the minimum bit energy (attained
at zero spectral efficiency) and the wideband slope which can be
achieved by pure OOK signaling.

\begin{appendix}

\section{Proof of Proposition \ref{prop:coherentoofskcap}} \label{appendix:coherentoofskcap}

Since the fading coefficients form a stationary ergodic process, the
capacity of OOFSK signaling can be formulated as follows:
\begin{gather*}
C(\ssnr) = \lim_{n \to \infty} \max_{X^n} \frac{1}{n} I(X^n;\R^n
\big| \, |h|^n ),
\end{gather*}
where $X^n = (X_1, \ldots, X_n)$, $\R^n = (\R_1, \ldots, \R_n)$, and
$|h|^n = (|h_1|, \ldots, |h_n|)$. As the additive Gaussian
noise samples 
are independent for each symbol interval, the conditional output
density satisfies
\begin{gather*}
p_{\R^n | \, X^n,|h|^n} = \prod_{k=1}^n p_{\R_k | \, X_k, |h_k|}
\end{gather*}
where
\begin{gather*}
p_{\R_k | \, X_k = i, |h_k|} = \left\{
\begin{array}{ll}
e^{-\sum_{j=1}^M R_{kj}} \, e^{-\alpha^2 |h_k|^2} \, I_0\left( 2
\sqrt{R_{ki} \, \alpha^2 |h_k|^2} \right) & 1 \le i \le M \\
e^{-\sum_{j=1}^M R_{kj}} & i = 0,
\end{array} \right.
\end{gather*}
with  $\alpha^2 = \frac{PT}{\nu N_0} = \frac{\ssnr}{\nu}$. From the
above fact, one can easily show that
\begin{align*}
I(X^n;\R^n \big| \, |h|^n ) &=  \sum_{k=1}^n I(X_k; \R_k \, \big| \,
|h_k|) - D\left( p_{\R^n | \, |h|^n} \, \bigg\| \, \prod_{k=1}^n
p_{\R_k | \, |h_k|} \bigg| \, F_{|h|^n} \right)
\\
& \le  \sum_{k=1}^n I(X_k; \R_k \, \big| \, |h_k|)
\end{align*}
where $D(\cdot||\cdot \big| \, F_{|h|^n})$ denotes the conditional
divergence. The above upper bound is achieved if the input vector
$X^n = (X_1, \ldots, X_n)$ has independent components. Due to the
symmetry of the channel, an input distribution equiprobable over
nonzero input values, i.e., $P(X_k = i) = \frac{\nu}{M}$ for $1 \le
i \le M$ where $P(X_k=0)=1-\nu$, maximizes $I(X_k; \R_k \, \big| \,
|h_k|)$ for each $k$. To see this, note that since the mutual
information is a concave function of the input vector, a sufficient
and necessary condition for an input vector to be optimal is
\begin{gather*}
\frac{\partial}{\partial P_i}\left[ I(X_k; \R_k \big| \, |h_k|) -
\lambda \left( \sum_{j=1}^M P_j - \nu \right) \right] = 0, \quad 1
\le i \le M
\end{gather*}
where $\lambda$ is a Lagrange multiplier for the equality constraint
$\sum_{j=1}^M P_j = \nu$, and $P_j$ denotes $P(X_k = j)$ for $1 \le
j \le M$. Note that the duty factor is fixed and hence $P(X = 0) = 1
- \nu$ is a predetermined constant. Evaluating the derivatives, the
above condition can be reduced to
\begin{gather*}
E_{|h_k|} \left\{ \int p_{\R_k\big| X_k = i, |h_k|} \log
\frac{p_{\R_k\big| X_k = i, |h_k|}}{p_{\R_k \big|\, |h_k| }} \, \ud
\R_k \right\} - 1 = \lambda, \quad 1 \le i \le M
\end{gather*}
and due to the symmetry of the channel, letting $P_i = P(X_k = i) =
\frac{\nu}{M}$ for $1 \le i \le M$ satisfies the condition.
Therefore an independent and identically distributed (i.i.d.) input
sequence with the above distribution achieves the capacity. The
capacity expression in (\ref{eq:coherentoofskcap}) is easily
obtained by evaluating the mutual information achieved by the
optimal input, considering a generic symbol interval, and dropping
the time index $k$. \hfill $\square$

\section{Proof of Proposition \ref{cor:coherentcapMinf}}
\label{appendix:coherentcapMinf}

The method of proof follows primarily from \cite{Butman} where
martingale theory is used to establish a similar result for $M$-ary
FSK signaling over the noncoherent Gaussian channel. The capacity
expression in (\ref{eq:coherentoofskcap}) can be rewritten as
\begin{align}
&C_M^p(\ssnr) = \nu E_{|h|} \left\{\int  e^{-R -
\frac{\ssnr}{\nu}|h|^2} I_0\left( 2\sqrt{\frac{\ssnr}{\nu} \, |h|^2
R}\right) \log \frac{e^{-R - \frac{\ssnr}{\nu}|h|^2} I_0\left(
2\sqrt{\frac{\ssnr}{\nu} \, |h|^2 R}\right)}{e^{-R}} \, \ud
R\right\}\nonumber
\\
&-E_{|h|}\left\{ \int e^{-\sum_{i=1}^M R_i} \,\, \frac{S_M(\R)}{M}
\log \frac{S_M(\R)}{M} \, \ud \R \right\} \label{eq:capinproof}
\end{align}
where the first term on the right-hand side can be recognized as the
conditional divergence
\\
$D(p_{R |\,\tilde{x},|h|} \, \big\| \, p_{R | \tilde{x} = 0, |h|}
\big| \, F_{|h|}F_{\tilde{x}})$, and
\begin{gather*}
S_M(\R) = \sum_{i=1}^M \big(\nu f(R_i,|h|,\ssnr) + (1- \nu)\big)
\end{gather*}
is a sum of i.i.d. random variables. The following result is noted
in \cite{Butman}.
\begin{lemma:martingale} \label{lemma:martingale}
Let $X_1, X_2, \cdots$ be identically distributed random variables
having finite mean. Let $S_n = X_1 + \cdots + X_n$, and $\beta_n =
\beta(S_n, S_{n+1}, \cdots)$, the Borel field generated by $S_n,
S_{n+1},\cdots$. Then $\left\{\cdots, \frac{S_n}{n},
\frac{S_{n-1}}{n-1}, \cdots, \frac{S_1}{1} \right\}$ is a martingale
with respect to $\{\cdots, \beta_n, \beta_{n-1}, \cdots, \beta_1\}$.
Moreover, if $g$ is a function which is convex and continuous on a
convex set containing the range of $X_1$, and if $E\{|g(X_1)|\} <
\infty$, then $\{g\left(\frac{S_n}{n} \right)\}_{\infty}$ is a
submartingale.
\end{lemma:martingale}
\noindent From Lemma \ref{lemma:martingale}, we conclude that
\begin{gather*}
\chi_M = g\left( \frac{S_M(\R)}{M} \right) = \frac{S_M(\R)}{M} \log
\frac{S_M(\R)}{M}
\end{gather*}
is a submartingale, and hence from the martingale convergence
theorem \cite{Grimmett}, $\chi_M$ converges to a limit $\chi_\infty$
almost surely and in mean. Therefore $\lim_{M \to \infty}
E\{\chi_M\} = E\{\lim_{M \to \infty} \chi_M\} = E\{\chi_\infty\}$.
Note also that from the strong law of large numbers and continuity
of the function $g(x) = x \log x$,
\begin{align}
\lim_{M \to \infty} \chi_M &= \lim_{M \to \infty} g\left(
\frac{S_M(\R)}{M} \right) \nonumber
\\
&= g\left( \lim_{M \to \infty} \frac{S_M(\R)}{M}\right) \nonumber
\\
&= g\left( E_R\{\nu f(R,|h|,\ssnr) + (1- \nu)\}\right) \nonumber
\\
&= g\left( \int e^{-R} \, \big(\nu f(R,|h|,\ssnr) + (1- \nu)\big) \,
\ud R \right) \nonumber
\\
&= g(1) = 0. \nonumber
\end{align}
Hence, we conclude that $\lim_{M \to \infty}
E_\R\left\{\frac{S_M(\R)}{M}\log \frac{S_M(\R)}{M} \right\}=0$. The
first term on the right-hand side of (\ref{eq:capinproof}) does not
depend on $M$, and the second term can be expressed as $E_{|h|} E_\R
\left\{ \frac{S_M(\R)}{M}\log \frac{S_M(\R)}{M} \right\}$. The proof
is completed by showing that
\begin{gather*}
\lim_{M \to \infty} E_{|h|} E_\R \left\{ \frac{S_M(\R)}{M}\log
\frac{S_M(\R)}{M} \right\} = E_{|h|} \left\{ \lim_{M \to \infty}
E_\R \left\{ \frac{S_M(\R)}{M}\log \frac{S_M(\R)}{M} \right\}
\right\} = 0,
\end{gather*}
where the interchange of limit and expectation needs to be justified
by invoking the Dominated Convergence Theorem. Note that since
$\left\{ \frac{S_M(\R)}{M}\log \frac{S_M(\R)}{M} \right\}$ is a
submartingale, $$0 \le E_\R \left\{ \frac{S_M(\R)}{M} \log
\frac{S_M(\R)}{M}\right\} \le E_R \{ S_1(R) \log S_1(R)\} < \infty.
$$
By noting that $f(R,|h|,\ssnr)$ is an exponentially decreasing
function of $|h|$, it can be easily shown that $$\int E_R \{ S_1(R)
\log S_1(R)\} \, \ud F_{|h|} < \infty $$ for any distribution
function $F_{|h|}$ with $E\{|h|^2\} < \infty$. Therefore, the
Dominated Convergence Theorem applies using the integrable upper
bound $ E_R \{ S_1(R) \log S_1(R)\}$. \hfill $\square$

\section{Proof of Proposition \ref{prop:coherentoofskcapwithphase}}
\label{appendix:coherentoofskcapwithphase}

Similarly to the proof of Proposition \ref{prop:coherentoofskcap},
an i.i.d. input sequence achieves the capacity and due to the
symmetry of the channel, equiprobable FSK signals each having
uniformly distributed phases are optimal. Now, the maximum
input-output mutual information is
\begin{align*}
I(X,\theta;\Y \big | \, h) = E_h \Bigg\{(1-\nu) \int p_{\Y |X =
0,\theta} \log \frac{p_{\Y |X=0,\theta}}{p_{\Y | \, |h|}} \, \ud \Y
\, \frac{1}{2 \pi} \, \ud \theta \\ + \left.\nu \int p_{\Y|X =
1,\theta,|h|} \log \frac{p_{\Y |X=1,\theta,|h|}}{p_{\Y | \, |h| }}
\, \ud \Y \, \frac{1}{2 \pi} \, \ud \theta \right \}
\end{align*}
where
\begin{align*}
p_{\Y | X = i, \theta_i, h} = \left\{
\begin{array}{ll}
\frac{1}{\pi^{M-1}} \, e^{-\sum_{j \neq i} |Y_j|^2} \frac{1}{\pi}
\, e^{-|Y_i - \alpha h e^{j\theta_i} |^2} & 1 \le i \le M \\
\frac{1}{\pi^M} \, e^{-\sum_{j = 1}^M |Y_j|^2} & i = 0.
\end{array} \right.
\end{align*}
In the above formulation, $\alpha^2 = \frac{PT}{\nu N_0} =
\frac{\ssnr}{\nu}$. It can be easily seen that
\begin{gather*}
\int p_{\Y | X = i,\theta} \log p_{\Y|X = i,\theta} \, \ud \Y \,
\frac{1}{2\pi} \, \ud \theta = - \log(\pi e)^M, \qquad 0 \le i \le
M.
\end{gather*}
The capacity expression in (\ref{eq:coherentoofskcapwithphase}) is
then obtained by first integrating
\begin{gather*}
\int p_{\Y|X = 0,\theta} \log p_{\Y} \, \ud \Y \, \frac{1}{2 \pi} \,
\ud \theta \quad \text{and} \quad \int p_{\Y|X = 1,\theta} \log
p_{\Y} \, \ud \Y \, \frac{1}{2 \pi} \, \ud \theta
\end{gather*}
with respect to $\theta$ and then making a change of variables, $R_j
= |Y_j|^2$. \hfill $\square$

\end{appendix}

\newpage

\end{spacing}
\end{document}